\documentclass[conference]{IEEEtran}

\usepackage{cite}
\usepackage{amsmath,amssymb,amsfonts,amsthm}
\usepackage{subcaption}
\usepackage{graphicx}
\usepackage{textcomp}
\usepackage{xcolor}
\usepackage{enumitem}
\usepackage{tabularx,array}
\usepackage{lipsum}
\usepackage{flushend}
\usepackage{booktabs,makecell,multirow}
\usepackage{algorithm}
\usepackage{algpseudocode}
\newif\ifshowchanges
\showchangestrue

\ifshowchanges
  
\else
  
\fi

\begin{document}

\title{Unexpected Far-Near-Far Transition in \\Mobile Near Field Terahertz Communications}

\author{
\IEEEauthorblockN{Peng Zhang, Hanmei Yuan, Zhe Wang, Vitaly Petrov, and Emil Bj\"ornson}
\IEEEauthorblockA{
Department of Communication Systems, KTH Royal Institute of Technology, Sweden \\
Email: \{pezhang, hanmei, zhewang2, vitalyp, emilbjo\}@kth.se
}
}

\maketitle

\begin{abstract}
At THz frequencies, the radiative near-field distance can be sufficiently large to matter in real deployments. Existing near-field formulas are often understood in a simple way: as the link distance decreases, the propagation regime is expected to change only once, i.e., from far field to near field. This paper shows that this intuition can fail for an elevated access point with downward tilt serving a ground user moving along the ground. Along such a path, the link distance and the viewing angle change together, so the near-field to far-field transition may take place more than once, creating an unexpected far-near-far transition. In this paper, we derive analytical conditions for when this transition occurs for tilted ULA-to-point and UPA-to-point scenarios and compute the corresponding transition point(s) on the ground. Numerical results validate the analysis and further show that this behavior depends strongly on the deployment geometry and can also arise at lower frequencies.
\end{abstract}

\section{Introduction}
Terahertz (THz, $300$\,GHz--$3$\,THz) communications have attracted considerable interest due to the large available bandwidth and the possibility of realizing highly directive antenna arrays~\cite{jiang2024terahertz}. In such systems, typically operating in line-of-sight, the so-called ``radiative near-field zone'' can become sufficiently large to matter over practical link distances~\cite{petrov2023mobile}. The size of the near-field zone determines the range of separation distances between the nodes where conventional plane-wave modeling and far-field beam steering remain accurate (far field), and where distance-dependent wavefront effects must be taken into account (radiative near field)~\cite{bjornson2021primer}.

The derivation of the near-field distance---the boundary condition separating the radiative near field from the far field---has been extensively studied in the literature to date. Starting from the classical Fraunhofer criterion based on a phase-error threshold~\cite{balanis2015antenna}, prior works have derived near-field distances for different array architectures, such as uniform linear arrays (ULAs) and uniform planar arrays (UPAs)~\cite{lu2023near,petrov2023near}, and for direction-dependent settings, including off-boresight cases~\cite{monemi2024study,li2025near,zhang2026nearfield}. Alternative definitions based on power uniformity, beam-focusing, spatial multiplexing, and capacity have also been investigated in~\cite{lu2021does,bjornson2021primer,bohagen2009spherical,jiang2005spherical}, among other works.

Despite the diversity of different criteria and studies on this topic, one observation remains consistent across the majority of related work on near-field distance derivation -- an explicit or an implicit assumption that, if a user equipment (UE) moves \emph{toward} the access point (AP), the switch from far-field zone to near-field zone \emph{happens only once} over the UE trajectory. In other words, there is a borderline 3D separation distance $r_{\text{F}}$, where all the UE--AP separation distances $r > r_{\text{F}}$ can be classified as far field, while all shorter AP--UE distances $r \leq r_{\text{F}}$ belong to the near field zone. \emph{However, as we demonstrate and analyze in the present paper, this observation doesn't always hold in practical deployment configurations.}

\begin{figure}[t]
    \centering
    \includegraphics[width=\linewidth]{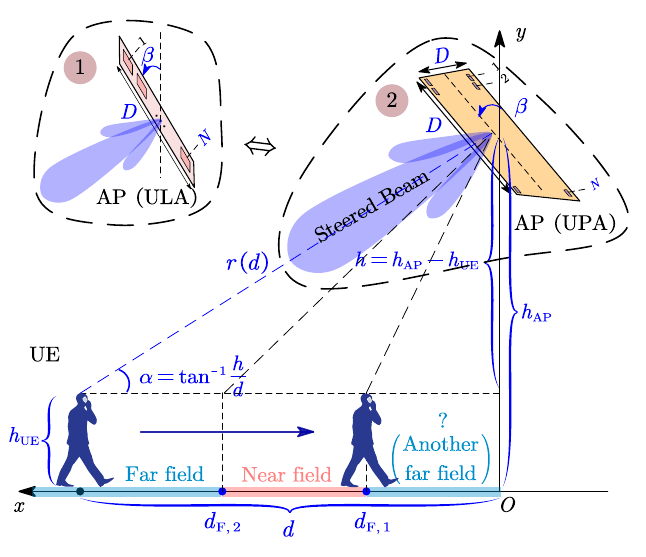}
    \caption{System model for the tilted ULA-to-point/UPA-to-point geometries. Case 1 is analyzed in Section~\ref{subsec:ula_point}, while Case 2 is analyzed in Section~\ref{subsec:upa_point}.}
    \vspace{-0.59cm}
    \label{fig:sys_model_combined}
\end{figure}

In practical deployments, instead, a UE often moves along the ground relative to an \emph{elevated and tilted} (e.g., downtilt) AP, as in Fig.~\ref{fig:sys_model_combined}. In such a configuration, the 3D link distance $r$ and the beam angle $\alpha$ vary simultaneously with the ground distance. As a result, since the near-field distance boundary condition is also angular-dependent~\cite{liu2024near,zhang2026nearfield,li2025near}, there could be \emph{more than one point along a straight UE trajectory}, where the conditions change from far field to near field and vice versa (see Fig.~\ref{fig:sys_model_combined}). This effect leads to a non-trivial \textit{far-near-far propagation transition} studied in this paper.

Motivated by the above-mentioned gap, this paper studies an array-to-point link and considers different AP antenna downtilt angles and height differences between the AP and the UE. We develop a mathematical framework for such a scenario and investigate how the propagation regime changes as the UE moves along the ground toward the AP. Then, for both ULA-to-point and UPA-to-point configurations, we analytically derive the conditions under which a far-near-far transition occurs and compute the corresponding transition point(s). Numerical results quantify how the transition depends on the deployment geometry and the carrier frequency, highlighting the setups where such a non-binary transition may and may not occur.

\section{System Model}\label{sec:system_model}
We consider a downlink scenario in which the UE is, for simplicity, modeled as an electrically small receiver. As illustrated in Fig.~\ref{fig:sys_model_combined}, $O$ is the origin of the ground coordinate system. The AP is located at height $h_{\mathrm{AP}}$, and the UE is located at height $h_{\mathrm{UE}} \leq h_{\mathrm{AP}}$. Let $d \ge 0$ denote the 2D horizontal ground distance of the UE from $O$. The corresponding 3D AP-UE link distance and beam angle are denoted by $r(d)$ and $\alpha$, respectively. Two AP array configurations are considered. In Case~1, the AP is equipped with a tilted ULA with aperture length $D$ and $N$ antenna elements. In Case~2, the AP is equipped with a tilted UPA with side length $D$ and $N \times N$ antenna elements. In both cases, the AP has a fixed downtilt by an angle $\beta \in [0,\pi/2)$ with respect to the vertical direction~\cite{chen2009antennas}.

\section{Analysis}
\label{sec:analysis}
In this section, we first analyze the ULA-to-point and UPA-to-point cases to characterize the propagation-regime transition, and then compute the corresponding transition point(s).

\subsection{ULA-to-Point Case}\label{subsec:ula_point}

We first consider Case~1 in Fig.~\ref{fig:sys_model_combined}, where the AP is equipped with a vertical $N \times 1$ ULA of length $D$. Using the phase-error-based near-field distance in~\cite{zhang2026nearfield}, where $\varphi$ denotes the phase-error threshold (e.g., $\varphi=\pi/8$ under the classical Fraunhofer criterion), the corresponding ULA near-field distance can be formulated as a function of the beam angle:
\begin{equation}
    r_{\mathrm F}(\alpha)
    =
    \frac{\pi D^2}{4 \lambda \varphi}\cos^2(\alpha-\beta).
    \label{eq:rF_ula}
\end{equation}
To determine the ground transition point(s) $d_{\text{F},1}$ and $d_{\text{F},2}$, we solve for the locations where the actual link distance equals the near-field distance. Using the geometric relation in Fig.~\ref{fig:sys_model_combined}, the link distance equals $r(\alpha)={h}/{\sin\alpha},$ where $h\triangleq h_{\mathrm{AP}}-h_{\mathrm{UE}}$ Therefore, the transition points satisfy $r(\alpha)=r_{\mathrm F}(\alpha)$, which can be rewritten~as
\begin{equation}
    \frac{4\lambda\varphi}{\pi D^2}h
    =
    \sin\alpha\,\cos^2(\alpha-\beta)\triangleq H_{\mathrm L}(\alpha,\beta).
    \label{eq:H_ula}
\end{equation}

The number of transition points is equal to the number of intersections between the horizontal line $\frac{4\lambda\varphi}{\pi D^2}h$ and the curve $H_{\mathrm L}(\alpha,\beta)$ over $\alpha\in[0,\pi/2]$. This intersection behavior is governed by the shape of $H_{\mathrm L}(\alpha,\beta)$, which can be characterized through its monotonicity and stationary points. In particular, its partial derivative with respect to $\alpha$ is
\begin{equation}
    {\partial H_{\mathrm L}(\alpha,\beta)}/{\partial \alpha}
    =
    \cos(\alpha-\beta)G(\alpha,\beta),
    \label{eq:H_ula_derivative}
\end{equation}
where
\begin{equation}
    G(\alpha,\beta)
    \triangleq
    \frac{3}{2}\cos(2\alpha-\beta)-\frac{1}{2}\cos\beta.
\label{eq:G_alpha}
\end{equation}

Since $\cos(\alpha-\beta)>0$ for $\alpha,\beta\in(0,\pi/2)$, the sign of $\partial H_{\mathrm L}(\alpha,\beta)/\partial\alpha$ is fully determined by the sign of $G(\alpha,\beta)$. Solving $G(\alpha,\beta)=0$ yields the unique stationary point
\begin{equation}
    \alpha_{\mathrm L}^\star
    =
    \frac{\beta}{2}
    +
    \frac{1}{2}
    \cos^{-1}\!\left(\frac{\cos\beta}{3}\right),
    \label{eq:alpha0_ula}
\end{equation}
which implies that, for any fixed $\beta$, $H_{\mathrm L}(\alpha,\beta)$ is increasing with $\alpha$ on $(0,\alpha_{\mathrm L}^\star)$ and decreasing with $\alpha$ on $(\alpha_{\mathrm L}^\star,\pi/2)$. Hence, for any fixed $\beta$, $H_{\mathrm L}(\alpha,\beta)$ attains its unique maximum over $\alpha\in(0,\pi/2)$ at
$\alpha=\alpha_{\mathrm L}^\star$, given by
\begin{equation}
    H_{\mathrm L}^{\max}(\beta)
    =
    H_{\mathrm L}(\alpha_{\mathrm L}^\star,\beta)
    =
    \sin\alpha_{\mathrm L}^\star\cos^2(\alpha_{\mathrm L}^\star-\beta).
    \label{eq:H_ula_max}
\end{equation}

We define the two ULA regime-transition height thresholds as
$h_{\mathrm{L},1}(\beta;\varphi)\triangleq \frac{\pi D^2}{4\lambda\varphi} H_{\mathrm L}\!\left(\frac{\pi}{2},\beta\right)=\frac{\pi D^2}{4\lambda\varphi}\sin^2\beta$
and
$h_{\mathrm{L},2}(\beta;\varphi)\triangleq \frac{\pi D^2}{4\lambda\varphi} H_{\mathrm L}^{\max}(\beta)$,
which divide the AP-height range into three regions corresponding to the near-to-far, far-near-far, and only-far propagation patterns.

Since $H_{\mathrm L}(0,\beta)=0$, $H_{\mathrm L}\!\left(\frac{\pi}{2},\beta\right)=\sin^2\beta$, and $H_{\mathrm L}(\alpha,\beta)$ attains its unique maximum $H_{\mathrm L}^{\max}(\beta)$ at $\alpha_{\mathrm L}^\star$, three cases may arise. Along the ground trajectory, we write $r(d)\triangleq r(\alpha(d))$ and $r_{\mathrm F}(d)\triangleq r_{\mathrm F}(\alpha(d))$, where $\alpha(d)=\tan^{-1}(h/d)$.

First, if $0 < h < h_{\mathrm{L},1}(\beta;\varphi)$, then the equation $r(\alpha)=r_{\mathrm F}(\alpha)$ admits a unique solution, denoted by $\alpha_2\in(0,\alpha_{\mathrm L}^\star)$. Let $d_{\mathrm F}=h\cot\alpha_2.$ Then, along the ground trajectory, we have $r(d)\le r_{\mathrm F}(d)$ for $0\le d\le d_{\mathrm F}$ and $r(d)>r_{\mathrm F}(d)$ for $d>d_{\mathrm F}$. Therefore, $d_{\mathrm F} = h\cot\alpha_2$ \textit{is the unique ground transition point}: the link is in the near-field regime for $0\le d\le d_{\mathrm F}$ and in the far-field regime for $d>d_{\mathrm F}$.

Second, if $h_{\mathrm{L},1}(\beta;\varphi)< h < h_{\mathrm{L},2}(\beta;\varphi)$, then the equation $r(\alpha)=r_{\mathrm F}(\alpha)$ has two distinct solutions, denoted by $\alpha_1>\alpha_2$. The resulting ground transition points are $d_{\mathrm F,1}=h\cot\alpha_1$ and $d_{\mathrm F,2}=h\cot\alpha_2$, which satisfy $d_{\mathrm F,1}<d_{\mathrm F,2}$. In this case,
\begin{equation}
\begin{cases}
r(d)>r_{\mathrm F}(d), & 0\le d<d_{\mathrm F,1} \qquad \text{(far field)},\\
r(d)\le r_{\mathrm F}(d), & d_{\mathrm F,1}\le d\le d_{\mathrm F,2} \qquad \text{(near field)},\\
r(d)>r_{\mathrm F}(d), & d>d_{\mathrm F,2} \qquad \text{(far field)}.
\end{cases}
\end{equation}

Therefore, as $d$ increases, the link undergoes an \textit{unexpected far-near-far transition}.\footnote{Note that the 3D link distance $r(d)=\sqrt{d^2+h^2}$ \emph{increases monotonically} with $d$. Hence, this far-near-far transition is caused \emph{not} by an inconsistent 3D vs. 2D distance evolution (if the UE would e.g., move closer to the AP in 2D plane but further away in the 3D plane), but rather by the joint variation of the 3D link distance, $r$, and the angle-dependent near-field distance, $r_{\text{F}}$.}

Finally, if $h > h_{\mathrm{L},2}(\beta;\varphi)$, then the equation $r(\alpha)=r_{\mathrm F}(\alpha)$ has no solution over $\alpha\in(0,\pi/2)$. Equivalently, $r(d)>r_{\mathrm F}(d)$ for all $d\ge 0$, which means that the link remains in the far-field regime for all UE locations on this trajectory.

Based on the above case-by-case analysis, \textbf{the propagation-regime transition along the ground can be classified as}
\begin{equation}
\scalebox{0.9}{$
\mathrm{Pattern}(h)=
\begin{cases}
\text{near-to-far}, &
0<h\le h_{\mathrm{L},1}(\beta;\varphi),\\[3pt]
\text{far-near-far}, &
h_{\mathrm{L},1}(\beta;\varphi)
< h \le
h_{\mathrm{L},2}(\beta;\varphi),\\[3pt]
\text{only-far}, &
h>h_{\mathrm{L},2}(\beta;\varphi).
\end{cases}$}
\label{eq:ula_regime_classification}
\end{equation}

Under the conventional $\varphi=\pi/8$ phase difference criterion~\cite{balanis2015antenna}, the two thresholds reduce to $h_{\mathrm{L},1}(\beta)=\frac{2D^2}{\lambda}\sin^2\beta$ and $        h_{\mathrm{L},2}(\beta)=\frac{2D^2}{\lambda}H_{\mathrm L}^{\max}(\beta),$ where $H_{\mathrm L}^{\max}(\beta)$ is given in \eqref{eq:H_ula_max}.

For a fixed UE height $h_{\mathrm{UE}}$, the above classification can be equivalently rewritten in terms of the AP height as
\begin{equation}
\scalebox{0.9}{$
\mathrm{Pattern}(h_{\mathrm{AP}})=
\begin{cases}
\text{near-to-far}, &
h_{\mathrm{UE}}<h_{\mathrm{AP}}\le h_{\mathrm{L},1}^{\mathrm{AP}}(\beta),\\[3pt]
\text{far-near-far}, &
h_{\mathrm{L},1}^{\mathrm{AP}}(\beta)
< h_{\mathrm{AP}} \le
h_{\mathrm{L},2}^{\mathrm{AP}}(\beta),\\[3pt]
\text{only-far}, &
h_{\mathrm{AP}}>h_{\mathrm{L},2}^{\mathrm{AP}}(\beta),
\end{cases}$}
\label{eq:ula_regime_classification_AP}
\end{equation}
where $h_{\mathrm{L},1}^{\mathrm{AP}}(\beta)\triangleq h_{\mathrm{UE}} + h_{\mathrm{L},1}(\beta)$ and $h_{\mathrm{L},2}^{\mathrm{AP}}(\beta)\triangleq h_{\mathrm{UE}} + h_{\mathrm{L},2}(\beta)$.

\subsection{UPA-to-Point Case}
\label{subsec:upa_point}

We now proceed with analyzing Case~2 (UPA-to-point geometry) in Fig.~\ref{fig:sys_model_combined}. The analysis, in general, follows similar steps to the one above for the ULA case, but with the UPA near-field distance given by a more complex equation~\cite{zhang2026nearfield}:
\begin{equation}
    r_{\mathrm F}(\alpha)
    =
    \frac{\pi D^2}{4\lambda\varphi}\Bigl[1+\cos^2(\alpha-\beta)\Bigr].
    \label{eq:rF_upa}
\end{equation}

Using the same geometric relation as in the ULA case, the transition condition $r(\alpha)=r_{\mathrm F}(\alpha)$ can be rewritten as
\begin{equation}
    \frac{4\lambda\varphi}{\pi D^2}h
    =
    H_{\mathrm P}(\alpha,\beta)
    \triangleq
    \sin\alpha\Bigl[1+\cos^2(\alpha-\beta)\Bigr].
    \label{eq:H_upa}
\end{equation}

Similar to the ULA case, the number of transition points is determined by the shape of $H_{\mathrm P}(\alpha,\beta)$ over $\alpha\in(0,\pi/2)$. Differentiating \eqref{eq:H_upa} with respect to $\alpha$ gives
\begin{equation}
    \frac{\partial H_{\mathrm P}(\alpha,\beta)}{\partial\alpha}
    =
    \cos\alpha\Bigl[1+\cos^2(\alpha-\beta)\Bigr]
    -
    \sin\alpha\sin\bigl(2(\alpha-\beta)\bigr).
    \label{eq:H_upa_derivative}
\end{equation}

The stationary condition $\partial H_{\mathrm P}(\alpha,\beta)/\partial\alpha=0$ becomes
\begin{equation}
    \tan\alpha
    =
    \frac{1+\cos^2(\alpha-\beta)}{\sin\bigl(2(\alpha-\beta)\bigr)}.
    \label{eq:stationary_upa}
\end{equation}

Let $t=\tan (\alpha-\beta)$ and $b=\tan\beta$, then \eqref{eq:stationary_upa} is equivalent~to $F_b(t)\triangleq bt^3+t^2+4bt-2=0.$ Since $\alpha\in(0,\pi/2)$, the feasible domain is $t\in(-b,\cot\beta)$. Over this interval, $F_b'(t)=3bt^2+2t+4b>0,$ and hence $F_b(t)$ is strictly increasing. Moreover, $F_b(-b)<0$ and $F_b(\cot\beta)>0$, which implies that $F_b(t)$ admits a unique root, denoted by $t^\star$. Therefore, $H_{\mathrm P}(\alpha,\beta)$ has a unique maximizer $\alpha_{\mathrm P}^\star=\beta+\arctan t^\star,$ with the maximum value
\begin{equation}
    \label{eq:Hmax_upa_general}
\scalebox{1.0}{$H_{\mathrm P}^{\max}(\beta)
    =
    H_{\mathrm P}\left(\alpha_{\mathrm P}^\star,\beta\right)
    =
    \frac{\bigl(t^\star\cos\beta+\sin\beta\bigr)\left((t^\star)^2+2\right)}
    {\left(1+(t^\star)^2\right)^{3/2}}.$}
\end{equation}

As in the ULA case, we define the two UPA regime-transition height thresholds as $h_{\mathrm{P},1}(\beta;\varphi)\triangleq \frac{\pi D^2}{4\lambda\varphi} H_{\mathrm P}\!\left(\frac{\pi}{2},\beta\right)=\frac{\pi D^2}{4\lambda\varphi}\bigl(1+\sin^2\beta\bigr)$ and $h_{\mathrm{P},2}(\beta;\varphi)\triangleq \frac{\pi D^2}{4\lambda\varphi} H_{\mathrm P}^{\max}(\beta)$. 

Since $H_{\mathrm P}(0,\beta)=0$, $H_{\mathrm P}(\pi/2,\beta)=1+\sin^2\beta$, as in the ULA case, one, two,  or no intersection may occur, which correspond to the near-to-far, far-near-far, and only-far propagation patterns along the ground, respectively. Hence, \textbf{the propagation-regime transition can be classified as}
\begin{equation}
\scalebox{0.90}{$
\mathrm{Pattern}(h)=
\begin{cases}
\text{near-to-far}, &
0<h\le h_{\mathrm{P},1}(\beta;\varphi),\\[3pt]
\text{far-near-far}, &
h_{\mathrm{P},1}(\beta;\varphi)
< h \le
h_{\mathrm{P},2}(\beta;\varphi),\\[3pt]
\text{only-far}, &
h>h_{\mathrm{P},2}(\beta;\varphi).
\end{cases}$}
\label{eq:upa_regime_classification}
\end{equation}

Similarly, for the conventional choice $\varphi=\pi/8$, the two thresholds reduce to $h_{\mathrm{P},1}(\beta)=\frac{2D^2}{\lambda}\bigl(1+\sin^2\beta\bigr)$ and $h_{\mathrm{P},2}(\beta)=\frac{2D^2}{\lambda}H_{\mathrm P}^{\max}(\beta),$ where $H_{\mathrm P}^{\max}(\beta)$ is given in \eqref{eq:Hmax_upa_general}.

For a fixed UE height $h_{\mathrm{UE}}$, the above classification can be equivalently rewritten in terms of the AP height as
\begin{equation}
\scalebox{0.89}{$
\mathrm{Pattern}(h_{\mathrm{AP}})=
\begin{cases}
\text{near-to-far}, &
h_{\mathrm{UE}}<h_{\mathrm{AP}}\le h_{\mathrm{P},1}^{\mathrm{AP}}(\beta),\\[3pt]
\text{far-near-far}, &
h_{\mathrm{P},1}^{\mathrm{AP}}(\beta)
< h_{\mathrm{AP}} \le
h_{\mathrm{P},2}^{\mathrm{AP}}(\beta),\\[3pt]
\text{only-far}, &
h_{\mathrm{AP}}>h_{\mathrm{P},2}^{\mathrm{AP}}(\beta),
\end{cases}$}
\label{eq:upa_regime_classification_AP}
\end{equation}
where $h_{\mathrm{P},1}^{\mathrm{AP}}(\beta)\!\triangleq h_{\mathrm{UE}} + h_{\mathrm{P},1}(\beta)$ and $h_{\mathrm{P},2}^{\mathrm{AP}}(\beta)\!\triangleq h_{\mathrm{UE}} + h_{\mathrm{P},2}(\beta)$.

\subsection{Computation of Transition Points}

Once the propagation-regime transition pattern is identified, the corresponding 2D ground AP--UE distance transition point(s) can be obtained by first solving the intersection equation in $\alpha$ and then mapping the resulting angle solution(s) to the ground distance via $d=h\cot\alpha$. To present both the ULA-to-point and UPA-to-point cases in a unified form, define $\Psi(\alpha)
    \triangleq
    H(\alpha,\beta)-\frac{4\lambda\varphi}{\pi D^2}h,$ where $H(\alpha,\beta)$ denotes $H_{\mathrm L}(\alpha,\beta)$ for the ULA case and $H_{\mathrm P}(\alpha,\beta)$ for the UPA case. The transition points are then determined by the roots of $\Psi(\alpha)=0$. Although neither \eqref{eq:H_ula} nor \eqref{eq:H_upa} admits a closed-form solution in general, $H(\alpha,\beta)$ is unimodal over $\alpha\in(0,\pi/2)$ and attains its unique maximum at $\alpha^\star$, where $\alpha^\star=\alpha_{\mathrm L}^\star$ for the ULA case and $\alpha^\star=\alpha_{\mathrm P}^\star$ for the UPA case. Hence, $\Psi(\alpha)$ is strictly increasing on $(0,\alpha^\star)$ and strictly decreasing on $(\alpha^\star,\pi/2)$, so its root(s) can be computed by bisection on the corresponding interval(s).

For the \textit{near-to-far} case, $\Psi(0)<0$ and $\Psi(\alpha^\star)>0$, so $\Psi(\alpha)=0$ has a unique root $\alpha_2\in(0,\alpha^\star)$, yielding the transition point $d_{\mathrm F}=h\cot\alpha_2$.

For the \textit{far-near-far} case, $\Psi(0)<0$, $\Psi(\alpha^\star)>0$, and $\Psi(\pi/2)<0$, so $\Psi(\alpha)=0$ has two roots: $\alpha_2\in(0,\alpha^\star)$ and $\alpha_1\in(\alpha^\star,\pi/2)$. The corresponding transition points are $d_{\mathrm F,1}=h\cot\alpha_1$ and $d_{\mathrm F,2}=h\cot\alpha_2$, with $d_{\mathrm F,1}<d_{\mathrm F,2}$.

For the \textit{only-far} case, $\Psi(\alpha)<0$ for all $\alpha\in(0,\pi/2)$, and thus no ground transition point exists.

\section{Numerical Results}\label{sec:simulation}

\begin{figure*}[!t]
    \centering
    \begin{minipage}{0.49\linewidth}
        \centering
        \includegraphics[width=\linewidth]{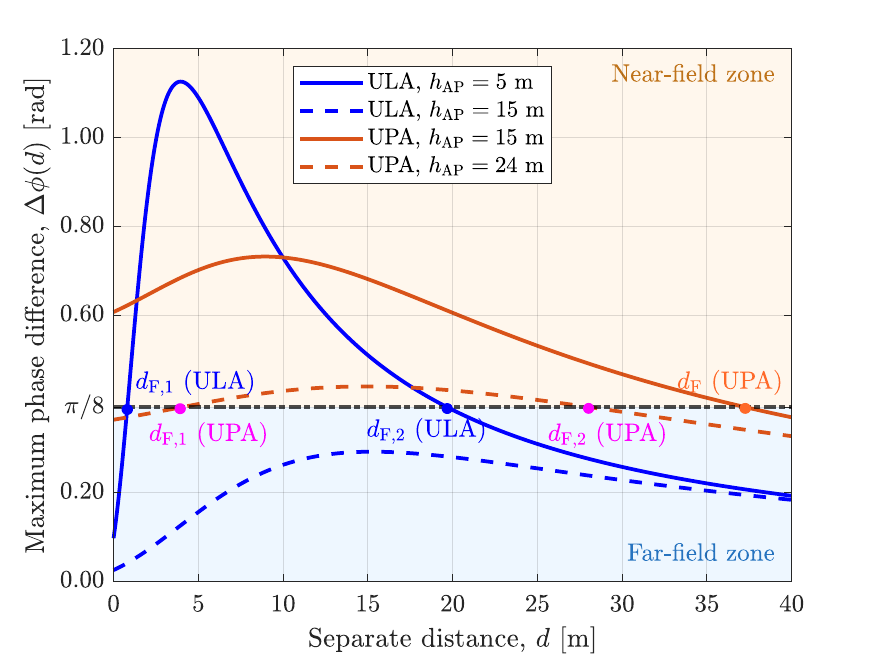}
        \caption{Maximum phase difference versus ground distance $d$ for representative ULA-to-point and UPA-to-point cases.}
        \label{fig:phase_difference_examples}
    \end{minipage}
    \hfill
    \begin{minipage}{0.49\linewidth}
        \centering
        \includegraphics[width=\linewidth]{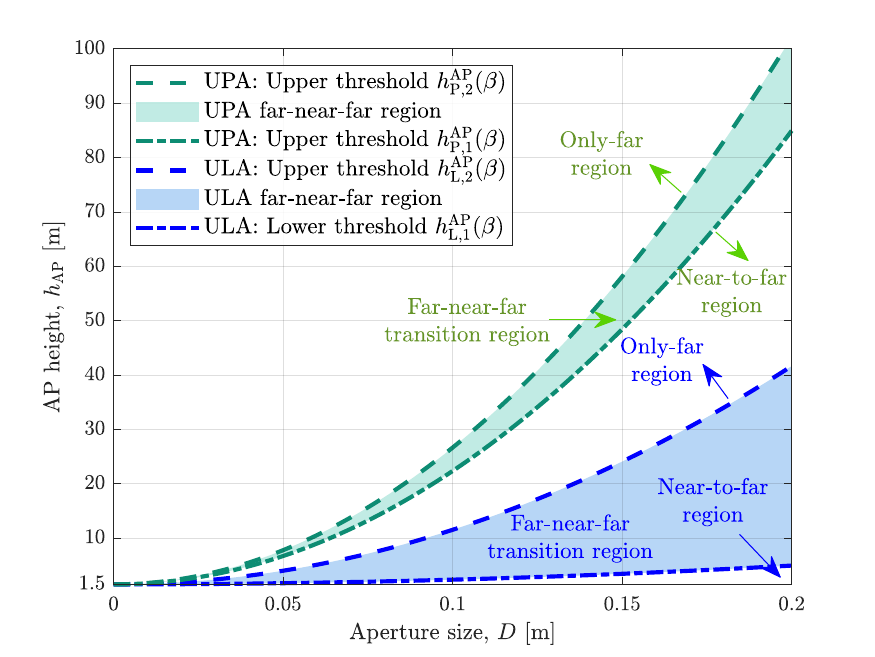}
        \caption{Comparison of the propagation-regime transition for ULA-to-point and UPA-to-point cases as a function of the aperture length $D$.}
        \label{fig:transition_regions}
    \end{minipage}
    \vspace{-5mm}
\end{figure*}

\begin{figure*}[!t]
    \centering
    \subfloat[ULA-to-point case\label{fig:ula_transition_region}]{
        \includegraphics[width=0.49\textwidth]{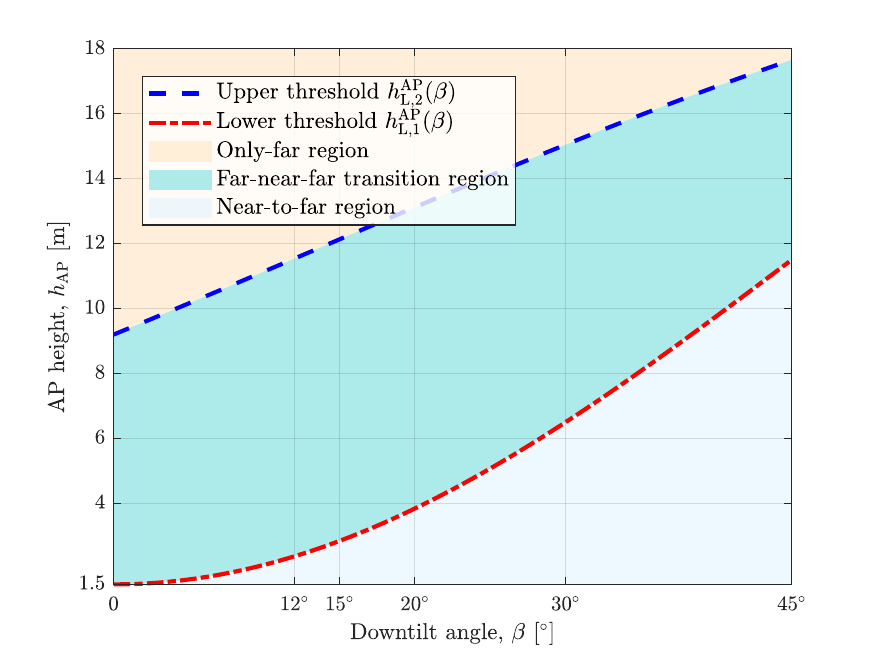}
    }
    \hspace{-4mm}
    \subfloat[UPA-to-point case\label{fig:upa_transition_region}]{
        \includegraphics[width=0.49\textwidth]{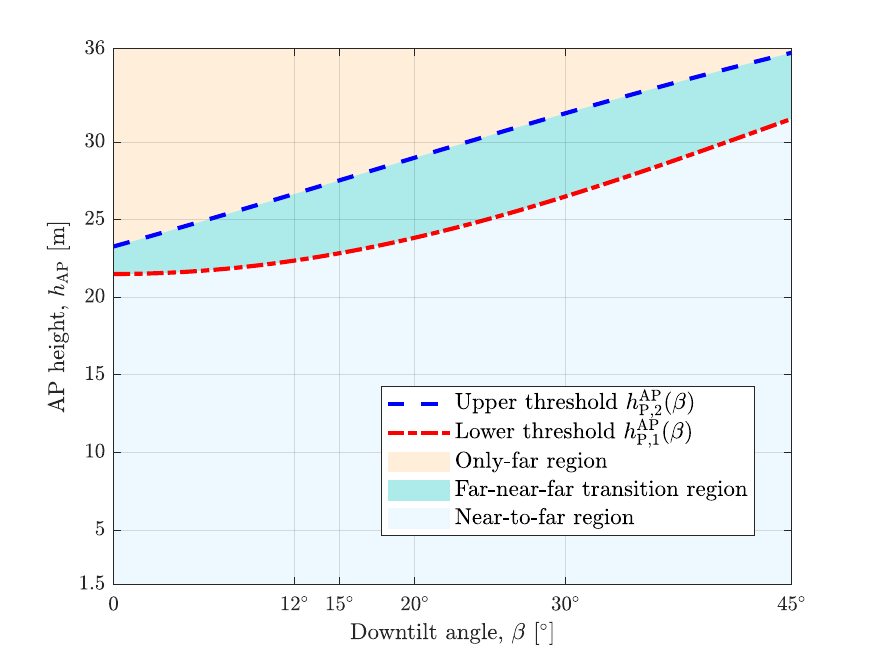}
    }
    \caption{Propagation-regime transition patterns for the ULA-to-point and UPA-to-point cases.}
    \label{fig:transition_region}
    \vspace{-5mm}
\end{figure*}

In this section, the analytical results for the far-near-far transition are numerically elaborated. Unless otherwise specified, we set the UE height to $h_{\mathrm{UE}}=1.5$ m~\cite{3gpp38901}, adopt the classical phase criterion $\varphi=\pi/8$, and use $f=300$ GHz (i.e., $\lambda=1$ mm) and $D=0.1$ m as the default system parameters.

We first illustrate the analytically predicted transition patterns through Fig.~\ref{fig:phase_difference_examples}, which plots the exact maximum phase difference along the ground trajectory. The value is computed using the exact geometry-dependent maximal phase-error expressions for the ULA-to-point and UPA-to-point cases~\cite{zhang2026nearfield}. Recall that differences above $\pi/8$ correspond to the radiative near field, whereas smaller values correspond to the far field. The figure shows ULA-to-point and UPA-to-point examples for a typical downtilt angle $\beta=12^\circ$~\cite{etsi_multipoint_downtilt}. The ULA case with $h_{\mathrm{AP}}=5$ m and the UPA case with $h_{\mathrm{AP}}=24$ m both exhibit two threshold crossings, corresponding to a far-near-far transition. By contrast, the UPA case with $h_{\mathrm{AP}}=15$ m shows a conventional near-to-far transition, whereas the ULA case with $h_{\mathrm{AP}}=15$ m remains in the far field over the shown range. These exact phase-difference curves therefore confirm all three analytically predicted transition patterns from Sec.~\ref{sec:analysis}.

We proceed with Fig.~\ref{fig:transition_regions}, exploring how the propagation-regime transition depends on the AP length $D$ and the AP height $h_{\mathrm{AP}}$. The far-near-far behavior appears only when $h_{\mathrm{AP}}$ lies between the two thresholds $h_{\mathrm{L},1}^{\mathrm{AP}}(\beta)/h_{\mathrm{P},1}^{\mathrm{AP}}(\beta)$ and $h_{\mathrm{L},2}^{\mathrm{AP}}(\beta)/h_{\mathrm{P},2}^{\mathrm{AP}}(\beta)$, while the near-to-far and only-far patterns occur below and above this interval, respectively. For a given aperture $D$, this far-near-far window is generally wider for the ULA case than for the UPA case, and it appears at lower, more practical AP heights. As $D$ increases, both thresholds move rapidly upward. This indicates that a larger aperture shifts the transition window to higher AP heights. Hence, the relevance of the far-near-far effect in THz systems depends not only on the array aperture, but also on the AP deployment height.

Fig.~\ref{fig:transition_region} shows how the downtilt angle $\beta$ affects the regime transition. Fig.~\ref{fig:ula_transition_region} shows how the downtilt angle $\beta$ affects the regime transition in the ULA-to-point case. The lower and upper height thresholds increase monotonically with $\beta$, which means that the far-near-far transition window shifts upward as the downtilt becomes larger. Fig.~\ref{fig:upa_transition_region} shows that the same behavior also appears in the UPA-to-point case, where the lower and upper height thresholds likewise increase monotonically with $\beta$. Therefore, increasing the downtilt does not change the three-region structure, but mainly changes where the transition occurs in terms of the AP height. This also implies that, for a fixed deployment height, varying $\beta$ may shift the operating point across different propagation regimes, e.g., from only-far to far-near-far, or further to near-to-far.

\begin{figure*}[!t]
    \centering
    \subfloat[Versus AP height $h_{\mathrm{AP}}$\label{fig:fig3_h}]{
        \includegraphics[width=0.48\textwidth]{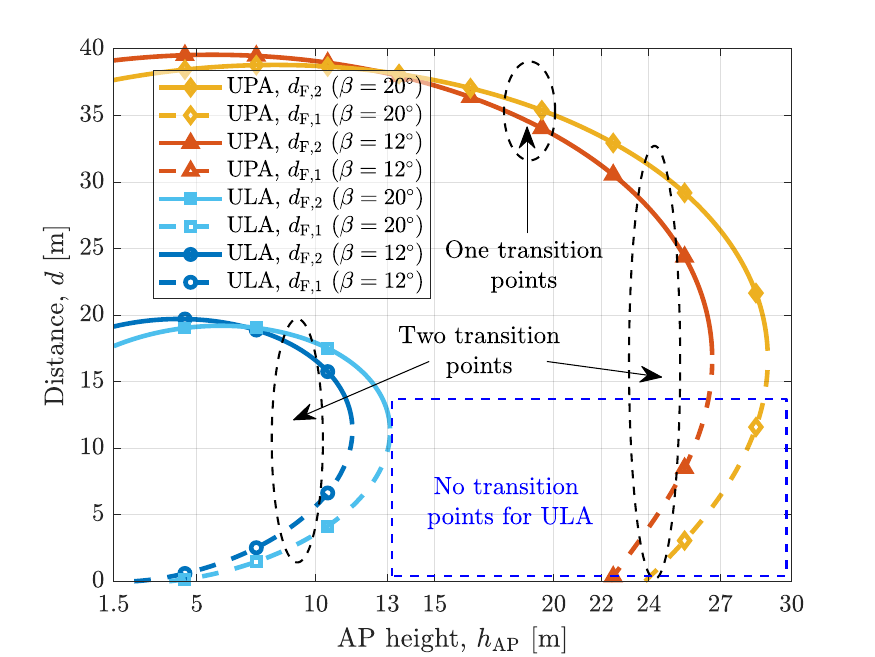}
    }
    \hfill
    \subfloat[Versus carrier frequency $f$\label{fig:transition_points_frequency}]{
        \includegraphics[width=0.48\textwidth]{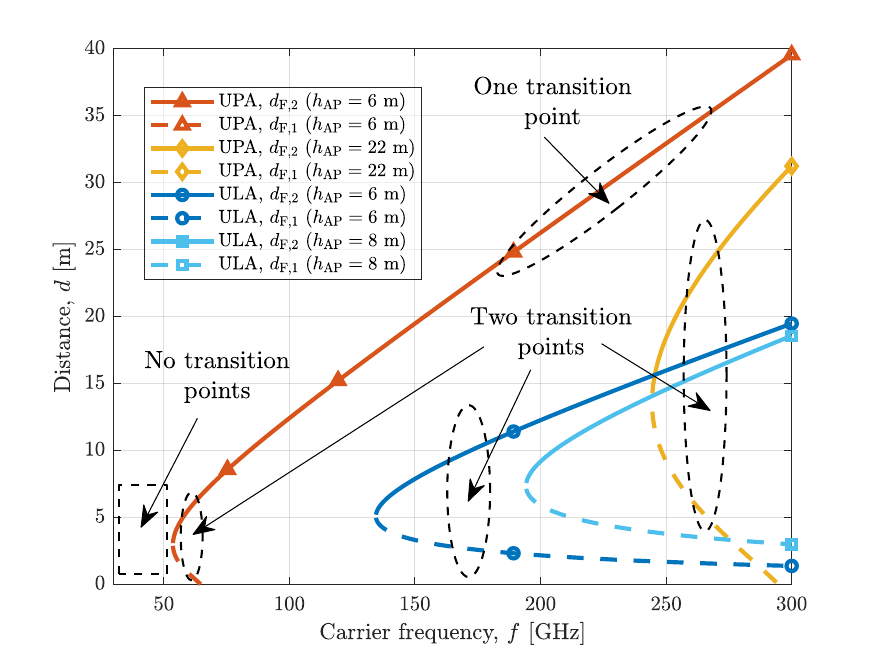}
    }
    \caption{Projected transition distance(s) for representative ULA-to-point and UPA-to-point cases versus (a) AP height $h_{\mathrm{AP}}$ and (b) carrier frequency $f$.}
    \label{fig:transition_points_combined}
\end{figure*}

Fig.~\ref{fig:transition_points_combined} further illustrates the projected transition distance(s) for representative ULA-to-point and UPA-to-point cases versus the AP height $h_{\mathrm{AP}}$ and the carrier frequency $f$. In Fig.~\ref{fig:fig3_h}, two representative downtilt angles, $\beta=12^\circ$ and $\beta=20^\circ$, are considered~\cite{etsi_multipoint_downtilt}. As predicted by the transition-region analysis, two projected transition distances, $d_{\mathrm F,1}$ and $d_{\mathrm F,2}$, coexist in the far-near-far regime. As $h_{\mathrm{AP}}$ increases, $d_{\mathrm F,1}$ moves outward while $d_{\mathrm F,2}$ moves inward, so that the near-field interval along the ground gradually shrinks and eventually disappears. Fig.~\ref{fig:transition_points_frequency} shows that this behavior is not limited to $300$ GHz: two projected transition points already exist over a broad frequency range below $300$ GHz in the shown examples. As $f$ increases, $d_{\mathrm F,1}$ moves toward the AP, whereas $d_{\mathrm F,2}$ extends to a larger 2D ground distance, which indicates an expansion of the near-field interval along the ground. Hence, the far-near-far behavior is not exclusive to the extreme THz regime, but \textit{may also arise at lower frequencies under certain geometric configurations}.

\section{Conclusion}\label{sec:conclusion}
In this paper, we characterized the propagation regime (far field to near field) transition along a practical UE ground trajectory in mobile THz communications with tilted AP arrays. For both ULA-to-point and UPA-to-point links, our analysis showed that, due to the coupled variation of the link distance and the beam angle, the propagation regime along the ground does not always change monotonically and \emph{may instead exhibit a complex far-near-far transition}. Numerical results further demonstrated that the existence and location of multiple far-field-to-near-field and near-field-to-far-field transition points depend strongly on the deployment geometry. These findings challenge the prevailing view that there is always a single separation distance between the far field and near field in mobile THz communications. The results also show that: (i)~the effect is more profound for ULA arrays than UPA arrays; and (ii)~this effect is not exclusively present in THz systems, but may also arise at lower frequencies (e.g., millimeter wave). The developed framework and the numerical results may facilitate more informed deployment choices in future millimeter wave/THz wireless communication systems, such as AP placement and downtilt configuration. 

\section*{Acknowledgment}
This work has been supported by the SweWIN center (Vinnova project 2023--00572), Digital Futures at KTH, Grant 2022--04222 from the Swedish Research Council, and the Swedish Foundation for Strategic Research FFL--9 program.

\bibliographystyle{IEEEtran}
\bibliography{references}

@article{zhang2026nearfield,
  title={Near-field Boundary Distance in mm{W}ave and {TH}z Communications with Misaligned Antenna Arrays},
  author={Zhang, Peng and Petrov, Vitaly and Bj{\"o}rnson, Emil},
  journal={arXiv preprint arXiv:2603.15311},
  year={2026}
}

@techreport{3gpp38901,
  author      = {{3rd Generation Partnership Project (3GPP)}},
  title       = {Study on channel model for frequencies from 0.5 to 100 {GH}z},
  institution = {},
  number      = {TR 38.901},
  version     = {19.2.0},
  year        = {2026}
}

@standard{etsi_multipoint_downtilt,
  author       = {{ETSI}},
  title        = {{EN 302 326-3 V1.3.1: Fixed Radio Systems; Multipoint Equipment and Antennas; Part 3: Parameters for Multipoint Antennas}},
  institution  = {European Telecommunications Standards Institute},
  year         = {2007}
}

@book{balanis2015antenna,
  title={Antenna theory: Analysis and design},
  author={Balanis, Constantine A},
  year={2015},
  publisher={Wiley}
}

@article{jiang2024terahertz,
  title={Terahertz communications and sensing for 6{G} and beyond: A comprehensive review},
  author={Jiang, Wei and others},
  journal={IEEE Commun. Surv. Tutor.},
  volume={26},
  number={4},
  pages={2326--2381},
  year={2024},
  month={Apr.}
}

@inproceedings{bjornson2021primer,
  title={A primer on near-field beamforming for arrays and reconfigurable intelligent surfaces},
  author={Bj{\"o}rnson, Emil and Demir, {\"O}zlem Tu{\u{g}}fe and Sanguinetti, Luca},
  booktitle={Proc. of the 55th Asilomar Conference},
  pages={105--112},
  year={2021},
  month={Oct.}
}

@article{liu2024near,
  title={Near-field communications: A comprehensive survey},
  author={Liu, Yuanwei and others},
  journal={IEEE Commun. Surv. Tut.},
  year={2025},
  month={Jun.}
}

@article{lu2023near,
  title={Near-field channel estimation in mixed {L}o{S}/{NL}o{S} environments for extremely large-scale {MIMO} systems},
  author={Lu, Yu and Dai, Linglong},
  journal={IEEE Trans. Commun.},
  volume={71},
  number={6},
  pages={3694--3707},
  year={2023},
  month={Jun.}
}

@inproceedings{petrov2023near,
  title={Near-field 6{G} Networks: Why Mobile Terahertz Communications MUST Operate in the Near Field},
  author={Petrov, Vitaly and Jornet, Josep Miquel and Singh, Arjun},
  booktitle={in Proc. of the IEEE GLOBECOM },
  pages={3983--3989},
  year={2023},
  month={Feb.}
}

@article{monemi2024study,
  title={A study on characterization of near-field sub-regions for phased-array antennas},
  author={Monemi, Mehdi and Bahrami, Sirous and Rasti, Mehdi and Latva-aho, Matti},
  journal={IEEE Trans. Commun.},
  volume={73},
  number={5},
  pages={2964--2979},
  year={2025},
  month={May}
}

@inproceedings{lu2021does,
  title={How does performance scale with antenna number for extremely large-scale {MIMO}?},
  author={Lu, Haiquan and Zeng, Yong},
  booktitle={Proc. of the IEEE ICC},
  year={2021},
  month={Aug.}
}

@article{bohagen2009spherical,
  title={On spherical vs. plane wave modeling of line-of-sight {MIMO} channels},
  author={Bohagen, Frode and Orten, Pal and Oien, Geir E},
  journal={IEEE Trans. Commun.},
  volume={57},
  number={3},
  pages={841--849},
  year={2009},
  month={Mar.}
}

@article{jiang2005spherical,
  title={Spherical-wave model for short-range {MIMO}},
  author={Jiang, Jeng-Shiann and others},
  journal={IEEE Trans. Commun.},
  volume={53},
  number={9},
  pages={1534--1541},
  year={2005},
  month={Sep.}
}

@article{li2025near,
  title={Near-field communications with extremely large-scale uniform arc arrays: Channel modeling and performance analysis},
  author={Li, Guoyu and You, Changsheng and Shang, Guanyu and Wu, Shaochuan},
  journal={IEEE Wireless Communications Letters},
  volume={14},
  number={4},
  pages={1009--1013},
  year={2025},
  month={Jan.}
}

@article{petrov2023mobile,
  author  = {V. Petrov and D. Bodet and A. Singh},
  title   = {Mobile near-field terahertz communications for {6G} and {7G} networks: {R}esearch challenges},
  journal = {Frontiers Commun. Netw.},
  volume  = {4},
  pages   = {1151324},
  year    = {2023},
  month   = {Mar.}
}

@book{chen2009antennas,
  title={Antennas for base stations in wireless communications},
  author={Chen, Zhi Ning and Luk, Kwai Man},
  year={2009},
  publisher={McGraw-Hill}
}
\end{document}